\documentclass[sigplan, screen, nonacm, a4paper]{acmart}
\pdfoutput=1
\usepackage{mathpartir}
\usepackage{cmll}
\usepackage{hyperref}

\author{Danielle Marshall}
\orcid{0000-0002-4284-3757}
\affiliation{
  \department{School of Computing}
  \institution{University of Kent}
  \country{}
}
\email{dm635@kent.ac.uk}        

\author{Dominic Orchard}
\orcid{0000-0002-7058-7842} 
\affiliation{
  \department{School of Computing}
  \institution{University of Kent}
  \country{}
}
 \email{D.A.Orchard@kent.ac.uk}

\usepackage{listings}
\lstdefinelanguage{Granule}{
  mathescape=true,
  morecomment=[l]{--},
  moredelim=[s][\itshape]{`}{`},
  showspaces=false,
  xleftmargin=2.5em,
  commentstyle=\itshape\color{black!60},
  basicstyle=\ttfamily\footnotesize,
  flexiblecolumns=true,
  columns=[l]flexible,
  columns=fullflexible,
  keepspaces=true,
  literate=
  {<}{\textcolor{effectColor}{<}}1
  {>}{\textcolor{effectColor}{>}}1
  {[}{\textcolor{coeffectColor}{[}}1
  {]}{\textcolor{coeffectColor}{]}}1
  {*}{\textcolor{uniqueColor}{*}}1
  {[r' : R']}{[{\textcolor{coeffectColor}{r' : R'}}]}1
  {[([}{\textcolor{coeffectColor}{[[}}2
  {])]}{\textcolor{coeffectColor}{]]}}2
  {!a}{\textcolor{coeffectColor}{!a}}1
  {!b}{\textcolor{coeffectColor}{!b}}1
  {*Coffee}{\textcolor{uniqueColor}{*Coffee}}1
  {*a}{\textcolor{uniqueColor}{*a}}1
  {*FloatArray}{\textcolor{uniqueColor}{*FloatArray}}1
  {forall}{$\forall$}1
  {Inf}{$\infty$}1
  {->}{$\rightarrow$}1
  {-o}{$\multimap$}1
  {=>}{$\Rightarrow$}1
  {<-}{\textcolor{effectColor}{$\leftarrow$}}1
  {/\\}{$\sqcap$}1
  {\\/}{$\sqcup$}1
  {<=}{$\leqslant$}1
  {>=}{$\geqslant$}1
  {\\}{$\lambda$}1
  {_1}{$\mathtt{_1}$}1
  {_2}{$\mathtt{_2}$}1
  {_3}{$\mathtt{_3}$}1
  {_4}{$\mathtt{_4}$}1
  {_L}{$\mathtt{_{L}}$}1
  {_LH}{$\mathtt{_{LH}}$}1
  {_Gr}{$\mathtt{_{Gr}}$}1
  {_p}{$\mathtt{_{p}}$}1
  {_q}{$\mathtt{_{q}}$}1
  {-o}{$\multimap$}1
  {\\times}{$\times$}1
  {--BLANK}{}1,
  keywordstyle = \color{blue!40!black},
  keywords = {data, type, let, in, case, of, if, then, else, where, import, Type, Semiring, Protocol},
  keywordstyle = [4]\bfseries\color{effectColor},
  keywords     = [4]{pure},
  numbers=left,
  numberstyle=\tiny\color{gray}
}

\lstnewenvironment{granule}
{\lstset{language=Granule}}{}

\lstnewenvironment{granuleInterface}
{\lstset{language=Granule,numbers=none}}{}

\newcommand{\granin}[1]{\lstinline[language=Granule]{#1}}

\usepackage[most]{tcolorbox}

\makeatletter
\newtcblisting[auto counter]{granulep}[1][]{%
  enhanced jigsaw,
  colframe=black!30,
  boxrule=0.5pt,
  toprule=0.5pt,
  bottomrule=0.5pt,
  top=0em,
  left=0em,
  bottom=0em,
  boxsep=0em,
  after skip=1em,
  before skip=1em,
  listing options={language={granule}},
  listing only,
  colback=white,
  sharp corners,
  overlay={
           \node[inner sep=2pt,xshift=-2cm,fill=white] (B) at (frame.south east) {Granule};
  },
  #1,
}
\makeatother

\makeatletter
\newtcblisting[auto counter]{granulepill}[1][]{%
  enhanced jigsaw,
  colframe=black!30,
  boxrule=0.5pt,
  toprule=0.5pt,
  bottomrule=0.5pt,
  top=0em,
  left=0em,
  bottom=0em,
  boxsep=0em,
  after skip=1em,
  before skip=1em,
  listing options={language={granule}},
  listing only,
  colback=white,
  sharp corners,
  overlay={
           \node[inner sep=2pt,xshift=-2cm,fill=white] (B) at (frame.south east) {Ill-typed Granule};
  },
  #1,
}
\makeatother
\definecolor{coeffectColor}{HTML}{0750d0}
\newcommand{\coeff}[1]{\textcolor{coeffectColor}{#1}}
\definecolor{uniqueColor}{HTML}{c20232}
\newcommand{\uniq}[1]{\textcolor{uniqueColor}{#1}}

\title[Graded Modal Types for Integrity and Confidentiality]{Graded Modal Types for\\ Integrity and Confidentiality}

\begin{abstract}
  Graded type systems, such as the one underlying the Granule programming
  language, allow various different properties of a program's behaviour to be
  tracked via annotating types with additional information, which we call
  \emph{grades}. One example of such a property, often used as a case study in
  prior work on graded types, is \emph{information flow control}, in which types
  are graded by a lattice of security levels allowing noninterference properties
  to be automatically verified and enforced. These typically focus on one
  particular aspect of security, however, known as \emph{confidentiality};
  public outputs are prohibited from depending on private inputs.
  \emph{Integrity}, a property specifying that trusted outputs must not depend
  on untrusted inputs, has not been examined in this context.

  This short paper aims to remedy this omission. It is well-known that
  confidentiality and integrity are in some sense dual properties, but simply
  reversing the ordering of the security lattice turns out to be unsatisfactory
  for the purpose of combining both kinds of property in a single system, at
  least in our setting. We analogize the situation to recent work on embedding
  both linear and uniqueness types in a graded framework, and use this framing
  to demonstrate that we can enforce both integrity and confidentiality
  alongside one another. The main idea is to add an additional flavour of
  modality annotated for integrity, such that the existing graded comonad for
  tracking confidentiality now also acts as a \emph{relative monad} over the new
  modality, with rules allowing information to flow from trusted to public to
  private.
\end{abstract}
\begin{document}
\maketitle
\section{Introduction and Motivation}
\label{sec:introduction}

Information flow control aims to track the flow of information through a program
when it is executed, to make sure that the program handles that information in a
secure way~\cite{1159651}. Secure information flow (discussed in the literature
since the 1970s~\cite{cohen1978information,10.1145/359636.359712}) encompasses
multiple aspects, with two of the most essential being \emph{confidentiality}
and \emph{integrity}. Pfleeger's textbook \emph{Security in
Computing}~\cite{10.5555/48805} describes these two properties as follows.
Confidentiality ``ensures that assets are accessed only by authorised parties'',
or in other words that private information is never accessed by a program which
only has public clearance. Meanwhile, integrity ``means that assets can be
modified only by authorised parties or only in authorised ways'', such that a
trusted program never depends on information from an untrusted source. The
strictest desirable property in both cases is \emph{noninterference}, which only
holds if public or trusted outputs may \emph{never} depend on private or
untrusted inputs respectively~\cite{6234468}, though this is often considered
difficult to achieve in practical systems.

Much of the prior work on using graded type systems for information flow control
aims to track and restrict the outputs that can be produced by a given
program~\cite{10.1145/292540.292555, 10.1145/268946.268976,
10.1145/596980.596983}. Implementations of such ideas also exist for more
widely-used functional languages such as Haskell~\cite{10.1145/2784731.2784756}.
More recently, graded type systems have been designed which enforce properties
based on \emph{coeffects}, where the inputs that can be passed into a program
are the focus. Such systems (with the type system underlying Granule being the
one we will focus on here~\cite{10.1145/3341714}) often make use of information
flow security as a case study in how annotating types with grades can allow more
properties of a program to be
verified~\cite{10.1145/2951913.2951939,10.1145/3408972,10.1145/3434331,10.1007/978-3-030-72019-3_17}.
However, these tend to only focus on confidentiality properties, which omits a
host of additional properties that could be guaranteed if it were also possible
to enforce integrity. Consider the following definition of a data type in
Granule.\footnote{The latest release of Granule is always available to download
and install from \url{https://github.com/granule-project/granule/releases}.}

\begin{granule}
  data Patient where
    Patient
      (Int    [Private]) -- Patient ID
      (String [Private]) -- Patient name
      (Int    [Public])  -- Patient age
\end{granule}

The type \granin{a [r]} means that we have a value of type \granin{a} wrapped
inside the $\square$ modality which must be used according to the restriction
described by the grade \granin{r}. Here, the patient's ID and name have the
grade \granin{Private}, while their age is \granin{Public}. Now, consider a
function with the type:

\begin{granule}
  meanAge : List Patient -> Int [Public]
\end{granule}

that calculates the mean age of a database (here simplified to a list) of
patients and returns the result at the public security level. Since ages are
also \granin{Public}, this is fine--but if we made a mistake while writing the
function and used private information such as IDs instead, this would be
rejected by the type checker due to the security annotations, so no information
can be leaked. Similar properties also apply when storing secret data
such as passwords or credit card numbers.

However, imagine a case where we are constructing a patient to add to the
database, as follows:

\begin{granule}
  addPatient : List Patient -> String [Private] 
            -> Int [Public] -> List Patient
\end{granule}

Here, we have security level grades restricting who will be able to view various
details once the database has been updated, but we have no way to stop some
compromised code from passing in an attempt at SQL injection using a string such
as \granin{"Alice'\); DROP TABLE patients;--"}, for example. If this input is
treated as trustworthy, we might well encounter dramatic problems later in our
program's execution.\footnote{\url{https://xkcd.com/327/}}

In order to avoid this, we would want some kind of grade that carries
information about the \emph{provenance} of our data~\cite{li2005unifying}, so we
could declare that we can only safely add patient information into our database
if the string verifiably comes from a trusted location. This would also be
useful if, for instance, we were encrypting private data and wanted a way to
ensure that our random numbers used for encryption were reliable and had not
been tainted by an untrusted source.

It is well understood that this kind of integrity property is dual in some sense
to the confidentiality properties which Granule can already
express~\cite{biba1977integrity} (though it is also known that this duality is
not sufficient to cover every facet of the concept of integrity, with more
complex mechanisms than a lattice model being required for some
applications~\cite{6234890, 10.1145/363516.363526}). It turns out that this
duality is closely comparable to a similar duality between \emph{linear} types
(forming the basis of Granule's graded type system) and \emph{uniqueness} types,
which has been more clearly elucidated in recent
work~\cite{10.1007/978-3-030-99336-8_13}. 

It turns out that while linear types are a restriction on what may happen in the
\emph{future} (a linear value must never be duplicated or discarded), uniqueness
types are a guarantee about what has happened in the \emph{past} (a unique value
must never \emph{have been} duplicated). We will now show how this understanding
also allows us to express integrity properties in Granule, through an additional
flavour of graded modality.

\section{Theory and Implementation}
\label{sec:details}

Our approach here builds on the type system described in the original Granule
paper~\cite{10.1145/3341714}, with some extra rules for the new modality
carrying integrity information.

The crucial insight here is that in order to combine confidentiality (public and
private) and integrity (trusted and untrusted) in a single system, we can treat
``public'' and ``untrusted'' as the \emph{same state}, both represented by the
$\coeff{\mathsf{Public}}$ grade; these both carry the same information, telling
us that there is no restriction on how the data may be used in the future (it is
not restricted to private usage) and we have no guarantee about how it was used
in the past (it did not necessarily come from a trusted source). The
$\coeff{\mathsf{Private}}$ grade behaves exactly as described in the original
work~\cite{10.1145/3341714}.

We introduce a new $\ast$ modality (whose syntax is borrowed from the
corresponding modality for uniqueness types~\cite{10.1007/978-3-030-99336-8_13},
as mentioned above) to carry the $\uniq{\mathsf{Trusted}}$ grade, with the
important rules for this modality's behaviour given below.
\begin{align*}
  &\inferrule*[right=\textsc{Reveal}]
  {\Gamma \vdash t : \ast_{\textcolor{uniqueColor}{\mathsf{Trusted}} }\ A}
  {\Gamma \vdash \text{reveal } t : \square_{\textcolor{coeffectColor}{\mathsf{Public}} }\ A}
  \quad\quad
  \inferrule*[right=\textsc{Nec}]
  {\varnothing \vdash t : A}
  {\varnothing \vdash \ast t : \ast_{\textcolor{uniqueColor}{\mathsf{Trusted}} }\ A}
  \\
  &\inferrule*[right=\textsc{Endorse}]
  {\Gamma_1 \vdash t_1 : \square_{\textcolor{coeffectColor}{\mathsf{Public}} }\ A \quad \Gamma_2, x : \ast_{\textcolor{uniqueColor}{\mathsf{Trusted}} }\ A \vdash t_2 : \square_{\textcolor{coeffectColor}{\mathsf{Public}} }\ B}
  {\Gamma_1 + \Gamma_2 \vdash \text{endorse } t_1 \text{ as } x \text{ in } t_2 :\square_{\textcolor{coeffectColor}{\mathsf{Public}} }\ B}
\end{align*}
The \textsc{Reveal} rule maps a trusted value to a public (untrusted) one,
allowing the information flow for integrity to behave as expected. The
\textsc{Endorse} rule allows a public value to be temporarily used as trusted
within the context of a larger computation; this mimics the common integrity
pattern of \emph{endorsement}, where a value is examined and declared to be
trusted to whatever extent is necessary for a particular usage~\cite{cite-key}.
The output, however, is required to be public, so that we can not leverage our
temporary integrity to `smuggle out' a trusted value outside the context of the
endorsement. These rules are accompanied by a necessitation rule, allowing
values to be trusted by default if they have no dependencies.

Note that the naming and pattern of the rules suggests a structural relationship
between the two modalities. The $\square$ modality previously acted as a
\emph{graded comonad}, but now when graded by $\coeff{\mathsf{Public}}$ also
acts as a \emph{relative monad}~\cite{10.1007/978-3-642-12032-9_21} over the
$\ast$ modality graded by $\uniq{\mathsf{Trusted}}$, with \textsc{Reveal} acting
as the `return' of the monad and \textsc{Endorse} acting as the `bind'; the
various required axioms for a relative monad hold here also.

The implementation follows much the same pattern. Granule already possesses a
\emph{semiring graded necessity modality}, where for a preordered semiring
$(\coeff{\mathcal{R}}, \coeff{*}, \coeff{1}, \coeff{+}, \coeff{0},
\coeff{\sqsubseteq})$, there is a family of types
$\{\square_{\textcolor{coeffectColor}{r} } A\}_{\coeff{r} \in
\coeff{\mathcal{R}}}$. (The standard example is the semiring of natural numbers,
where grades describe exactly how many times variables are used; $\coeff{1}$
captures \emph{linear} usage.) Granule's confidentiality tracking is based on
the semiring $\{\coeff{\mathsf{Public}},\ \coeff{\mathsf{Private}}\}$, where
$\coeff{0} = \coeff{\mathsf{Private}}$, $\coeff{1} = \coeff{\mathsf{Public}}$,
$\coeff{*} = \coeff{\vee}$ and $\coeff{+} = \coeff{\wedge}$. The implementation
provides an additional $\uniq{\ast}$ modality obeying the above rules along with
syntactic sugar for applying them. This allows us to repair the earlier example:

\begin{granule}
  addPatient : List Patient -> String *{Trusted}
            -> Int [Public] -> List Patient
\end{granule}

Now, only a string from a trusted source will be accepted as the name of a
patient added to our database. Note that we do not need to change the definition
of our data type, since the string can be converted to
$\coeff{\mathsf{Private}}$ through a combination of the \textsc{Return} rule and
the \emph{approximation} rule that already exists in Granule's type system,
which in this case allows $\coeff{\mathsf{Public}}$ information to be treated as
$\coeff{\mathsf{Private}}$ since $\coeff{\mathsf{Private}}\ \coeff{\sqsubseteq}\
\coeff{\mathsf{Public}}$ is given by the ordering of the semiring.

\section{Conclusion and Future Work}
\label{sec:future}

We have shown that simple integrity properties can be expressed in Granule
alongside confidentiality, leveraging a similar duality to one that
exists between linearity and uniqueness. Directions for further research could include:

\begin{itemize}
  \item Extending Granule's capacity to track confidentiality and integrity
  further, going beyond lattices containing only two security levels in order to
  enforce more complex and fine-grained properties.
  \item Considering additional aspects of information flow security such as
  \emph{availability}, where information which should always be available cannot
  depend on information that may be unavailable; the direction of information
  flow here is the same as for integrity, so this should be possible using the
  new flavour of modality.
  \item Borrowing the ideas currently being developed for extending Granule's
  uniqueness types to a more complex \emph{ownership} model via fractional
  grading on the $\ast$ modality~\cite{10.1007/3-540-44898-5_4, fractional},
  which here may allow us to express integrity properties relating to
  \emph{separation of duties}~\cite{li2005unifying}; this would be a starting
  point for capturing further aspects of integrity that go beyond what can be
  described by the current lattice-based model~\cite{6234890}.
  \item It would be interesting to explore how this idea connects to recent work
  on bridging the gap between monadic and comonadic approaches to information
  flow analyses~\cite{10.1145/3563335}. Note that Granule's current
  confidentiality tracking is primarily comonadic in nature, but also
  incorporates a touch of monadic flavour through use of a \granin{flatten}
  operation, which allows transformations such as
  $\square_{\textcolor{coeffectColor}{\mathsf{Public}}}
  (\square_{\textcolor{coeffectColor}{\mathsf{Private}}}\ A) \rightarrow
  \square_{\textcolor{coeffectColor}{\mathsf{Private}}}\ A$; this operation is
  derivable from Granule's rules for pattern matching on nested modalities~\cite{10.1145/3341714}.
\end{itemize}

This paper also forms part of a larger body of work that involves uncovering a
general algebraic structure underlying \emph{global guarantees} about a program,
which are best represented using a graded comonad that also acts as a relative
monad over some functor. This work is itself ongoing.

\bibliography{references}

\end{document}